# Large payload bidirectional quantum secure direct communication without information leakage

Tian-Yu Ye*

College of Information & Electronic Engineering, Zhejiang Gongshang University, Hangzhou 310018, P.R.China

**Abstract**

A large payload bidirectional quantum secure direct communication (BQSDC) protocol without information leakage is proposed, which is based on entanglement swapping between any two GHZ states. Two remote authorized parties, Alice and Bob, can safely exchange their individual secret messages without worrying about the information leakage problem. Our protocol uses a shared secret GHZ state to overcome the information leakage problem. The shared secret GHZ state plays two roles in the bidirectional communication process: on one hand, it lets Bob know the prepared initial state; on the other hand, it is used for encoding Bob's secret messages. Moreover, our protocol can transmit six bits of secret messages per round communication. Compared with those previous BQSDC protocols, the advantage of our protocol lies in having the following two characters simultaneously: on one hand, it overcomes the information leakage problem; one the other hand, its capacity is as high as six bits per round communication.

## 1 Introduction

Quantum communication has been one of the most important applications of quantum mechanics, and has attracted a lot of attention. As one branch of quantum communication, quantum key distribution (QKD) aims to establish a shared secret key with unconditional security between two remote authorized communication parties. Since the first QKD was proposed by Bennett and Brassard [1] in 1984, a lot good QKD protocols [2-4] have been put forward. Subsequently, the new concept of quantum secure direct communication (QSDC) was proposed, whose object is to send secret messages directly through quantum channel without establishing a key to encrypt them firstly. In 2002, a pioneering QSDC was proposed by Beige et al. [5], which still needs to transmit the additional classical information to read out the secret information taken by each qubit. Subsequently, Bostrom and Felbinger [6] suggested the famous Ping-Pong protocol using the idea of dense coding of Einstein-Podolsky-Rosen(EPR) pairs. In 2003, Deng et al. [7] put forward a two-step QSDC protocol based on Bell states. In 2004, Cai et al.[8] introduced two additional unitary operations into Ping-Pong protocol to enhance its capacity. In 2005, Wang et al.[9] proposed a multi-step QSDC protocol based on multi-particle Greenberger–Horne–Zeilinger (GHZ) states. In recent years, QSDC has progressed rapidly[10-14].

However, in 2004, Zhang et al.[15-17] and Nguyen[18] found out that a lot of previous QSDC protocols are message-unilaterally-transmitted communication protocols, and proposed the new concept of bidirectional quantum secure direct communication (BQSDC) to realize the bidirectional communication. Subsequently, Man et al. [19] found out that Nguyen' protocol[18] cannot resist the intercept-and-resend attack and presented an improved version accordingly. In 2006, Jin et al. [20] presented a three-party simultaneous QSDC based on a GHZ state; Man and Xia[21] presented a controlled BQSDC based on a GHZ state; Ji and Zhang[22] proposed a quantum dialogue protocol based on a single photon; Man et al.[23] proposed a quantum dialogue protocol based on entanglement swapping of GHZ states. In 2007, Man and Xia[24] found out that Jin's protocol[20] has the information definite leakage problem, and suggested a solution to solve this problem; Chen et al.[25] proposed a BQSDC based on entanglement swapping of Bell sates; Yang et al.[26] presented a quasi-secure quantum dialogue scheme based on a single photon. In 2008, Gao et al. [27] pointed out from the viewpoint of information theory and cryptography that both Jin's protocol[20] and Man's improved version[24] have the information leakage problem; Gao et al.[28] pointed out that all of Nguyen's protocol[18], Man's protocol[19], Ji's protocol[22] and Man's protocol[23] also have the information leakage problem; Tan and Cai[29] analyzed the information leakage problem in Nguyen' protocol[18] from the viewpoint of Holevo Bound. In 2009, Shan et al.[30] put forward a quantum dialogue protocol based on entanglement swapping of two Bell states via cavity Quantum electro-dynamics (QED); Shi et al.[31] proposed a BQSDC without information leakage based on a normal Bell state and an auxiliary Bell state. In 2010, Shi et al.[32] proposed a quantum dialogue protocol without information leakage based on a single photon and an auxiliary single photon; Shi[33] proposed a BQSDC using the correlation extractability of Bell state and an auxiliary particle to overcome the information leakage problem; Gao[34] proposed two quantum dialogue protocols without information leakage based on entanglement swapping between two Bell states. In 2013, Ye and Jiang[35] presented two approaches to solve the information definite leakage in Man's protocol [21]. However, as pointed out in Refs.[36-37],Ye's two improved protocols[35] still have the information leakage problem. Actually, all of the protocols in Refs.[15,16,25,26,30] also have the information leakage problem. As analyzed above, it can be concluded that the information leakage problem has been a great security threat to BQSDC, since a lot BQSDC protocols [15,16,18-26,30,35] have the drawback of information leakage. Moreover, a closer analysis can discover that the amount of secret messages exchanged through those above BQSDC protocols [15,16,18-26,30-35] is still not large enough.

*Corresponding author:
 E-mail：yetianyu@mail.zjgsu.edu.cn

In this paper, we propose a large payload BQSDC protocol without information leakage, based on entanglement swapping between any two GHZ states. Compared with those previous BQSDC protocols[15,16,18-26,30-35], the advantage of our protocol lies in having the following two characters simultaneously: on one hand, it overcomes the information leakage problem using a shared secret GHZ state; one the other hand, its capacity is as high as six bits per round communication.

## 2 Encoding scheme

As we know, the GHZ states are three-particle maximally entangled states, which can be listed as follows:

$$|\Psi_0\rangle = \frac{1}{\sqrt{2}}(|000\rangle+|111\rangle), |\Psi_1\rangle = \frac{1}{\sqrt{2}}(|000\rangle-|111\rangle), |\Psi_2\rangle = \frac{1}{\sqrt{2}}(|100\rangle+|011\rangle), |\Psi_3\rangle = \frac{1}{\sqrt{2}}(|100\rangle-|011\rangle),$$

$$|\Psi_4\rangle = \frac{1}{\sqrt{2}}(|010\rangle+|101\rangle), |\Psi_5\rangle = \frac{1}{\sqrt{2}}(|010\rangle-|101\rangle), |\Psi_6\rangle = \frac{1}{\sqrt{2}}(|110\rangle+|001\rangle), |\Psi_7\rangle = \frac{1}{\sqrt{2}}(|110\rangle-|001\rangle). \quad (1)$$

As pointed out by the dense coding scheme of GHZ states presented by Lee et al.[38], one GHZ state can be transformed into another after performed with single-particle unitary operations on its any two particles. Define the eight composite unitary operations as $U_0 = \sigma_z \otimes \sigma_z, U_1 = I \otimes \sigma_z, U_2 = i\sigma_y \otimes \sigma_z, U_3 = \sigma_x \otimes \sigma_z, U_4 = I \otimes \sigma_x, U_5 = \sigma_z \otimes \sigma_x, U_6 = \sigma_x \otimes \sigma_x$ and $U_7 = i\sigma_y \otimes \sigma_x$, where $I = |0\rangle\langle 0|+|1\rangle\langle 1|$, $\sigma_z = |0\rangle\langle 0|-|1\rangle\langle 1|, \sigma_x = |0\rangle\langle 1|+|1\rangle\langle 0|$ and $i\sigma_y = |0\rangle\langle 1|-|1\rangle\langle 0|$ are four single-particle unitary operations. Consequently, after performed with $U_k$ $(k=0,1,\cdots,7)$ on its first and second particles, one GHZ state will be transformed into another in the way shown in Table 1. Let each $U_k$ correspond to three bits information, i.e., $U_0 \to 000, U_1 \to 001, U_2 \to 010, U_3 \to 011, U_4 \to 100, U_5 \to 101, U_6 \to 110, U_7 \to 111$.

Table 1. The transformation relations between any two GHZ states
(The states in column denote the initial states, while the states in row denote the transformation outcomes)

|  | $|\Psi_0\rangle$ | $|\Psi_1\rangle$ | $|\Psi_2\rangle$ | $|\Psi_3\rangle$ | $|\Psi_4\rangle$ | $|\Psi_5\rangle$ | $|\Psi_6\rangle$ | $|\Psi_7\rangle$ |
|---|---|---|---|---|---|---|---|---|
| $|\Psi_0\rangle$ | $U_0$ | $U_1$ | $U_2$ | $U_3$ | $U_4$ | $U_5$ | $U_6$ | $U_7$ |
| $|\Psi_1\rangle$ | $U_1$ | $U_0$ | $U_3$ | $U_2$ | $U_5$ | $U_4$ | $U_7$ | $U_6$ |
| $|\Psi_2\rangle$ | $U_2$ | $U_3$ | $U_0$ | $U_1$ | $U_6$ | $U_7$ | $U_4$ | $U_5$ |
| $|\Psi_3\rangle$ | $U_3$ | $U_2$ | $U_1$ | $U_0$ | $U_7$ | $U_6$ | $U_5$ | $U_4$ |
| $|\Psi_4\rangle$ | $U_4$ | $U_5$ | $U_6$ | $U_7$ | $U_0$ | $U_1$ | $U_2$ | $U_3$ |
| $|\Psi_5\rangle$ | $U_5$ | $U_4$ | $U_7$ | $U_6$ | $U_1$ | $U_0$ | $U_3$ | $U_2$ |
| $|\Psi_6\rangle$ | $U_6$ | $U_7$ | $U_4$ | $U_5$ | $U_2$ | $U_3$ | $U_0$ | $U_1$ |
| $|\Psi_7\rangle$ | $U_7$ | $U_6$ | $U_5$ | $U_4$ | $U_3$ | $U_2$ | $U_1$ | $U_0$ |

Now let us consider the entanglement swapping between any two GHZ states. Without loss of generality, we take the entanglement swapping between two $|\Psi_0\rangle$'s shown in formula (2) for example. According to formula (2), if the Bell-basis measurements are performed on $A_1A_2$, $B_1B_2$ and $C_1C_2$, respectively, after entanglement swapping, two $|\Psi_0\rangle$'s composed of $A_1B_1C_1$ and $A_2B_2C_2$ will collapse to eight different outcome combinations of $A_1A_2$, $B_1B_2$ and $C_1C_2$ with equal probability. Here, the subscripts $A_i$, $B_i$ and $C_i$ ($i=1,2$) denote the three particles of a GHZ state, respectively.

$$|\Psi_0\rangle_{A_1B_1C_1} \otimes |\Psi_0\rangle_{A_2B_2C_2} = \left(\frac{1}{\sqrt{2}}\right)^3 [|\Phi^+\rangle_{A_1A_2}|\Phi^+\rangle_{B_1B_2}|\Phi^+\rangle_{C_1C_2} + |\Phi^+\rangle_{A_1A_2}|\Phi^-\rangle_{B_1B_2}|\Phi^-\rangle_{C_1C_2} + |\Phi^-\rangle_{A_1A_2}|\Phi^+\rangle_{B_1B_2}|\Phi^-\rangle_{C_1C_2}$$

$$+ |\Phi^-\rangle_{A_1A_2}|\Phi^-\rangle_{B_1B_2}|\Phi^+\rangle_{C_1C_2} + |\Psi^+\rangle_{A_1A_2}|\Psi^+\rangle_{B_1B_2}|\Psi^+\rangle_{C_1C_2} + |\Psi^+\rangle_{A_1A_2}|\Psi^-\rangle_{B_1B_2}|\Psi^-\rangle_{C_1C_2}$$

$$+ |\Psi^-\rangle_{A_1A_2}|\Psi^+\rangle_{B_1B_2}|\Psi^-\rangle_{C_1C_2} + |\Psi^-\rangle_{A_1A_2}|\Psi^-\rangle_{B_1B_2}|\Psi^+\rangle_{C_1C_2}] \quad (2)$$

where $|\Phi^\pm\rangle = \frac{1}{\sqrt{2}}(|00\rangle\pm|11\rangle)$ and $|\Psi^\pm\rangle = \frac{1}{\sqrt{2}}(|01\rangle\pm|10\rangle)$ are four Bell states. Let us further code the different outcome combinations of $A_1A_2$, $B_1B_2$ and $C_1C_2$ from formula (2) into the outcome collection as shown in formula (3).

$$\{|\Phi^+\rangle_{A_1A_2}|\Phi^+\rangle_{B_1B_2}|\Phi^+\rangle_{C_1C_2}, |\Phi^+\rangle_{A_1A_2}|\Phi^-\rangle_{B_1B_2}|\Phi^-\rangle_{C_1C_2}, |\Phi^-\rangle_{A_1A_2}|\Phi^+\rangle_{B_1B_2}|\Phi^-\rangle_{C_1C_2}, |\Phi^-\rangle_{A_1A_2}|\Phi^-\rangle_{B_1B_2}|\Phi^+\rangle_{C_1C_2},$$
$$|\Psi^+\rangle_{A_1A_2}|\Psi^+\rangle_{B_1B_2}|\Psi^+\rangle_{C_1C_2}, |\Psi^+\rangle_{A_1A_2}|\Psi^-\rangle_{B_1B_2}|\Psi^-\rangle_{C_1C_2}, |\Psi^-\rangle_{A_1A_2}|\Psi^+\rangle_{B_1B_2}|\Psi^-\rangle_{C_1C_2}, |\Psi^-\rangle_{A_1A_2}|\Psi^-\rangle_{B_1B_2}|\Psi^+\rangle_{C_1C_2}\} \to \mathbb{C}_0 \quad (3)$$

Similarly, just as pointed out in Ref.[39], after further extending the second GHZ state in the initial state group of formula(2) from $|\Psi_0\rangle$ to other seven GHZ states, there are other seven corresponding outcome collections composed of the different outcome combinations of $A_1A_2$, $B_1B_2$ and $C_1C_2$, which can be shown as formulas (4)-(10). According to formulas (3)-(10), after entanglement swapping, each outcome combination of $A_1A_2$, $B_1B_2$ and $C_1C_2$ does only correspond to one initial state group of $A_1B_1C_1$ and $A_2B_2C_2$.

$$\{|\Phi^+\rangle_{A_1A_2}|\Phi^+\rangle_{B_1B_2}|\Phi^-\rangle_{C_1C_2}, |\Phi^+\rangle_{A_1A_2}|\Phi^-\rangle_{B_1B_2}|\Phi^+\rangle_{C_1C_2}, |\Phi^-\rangle_{A_1A_2}|\Phi^+\rangle_{B_1B_2}|\Phi^+\rangle_{C_1C_2}, |\Phi^-\rangle_{A_1A_2}|\Phi^-\rangle_{B_1B_2}|\Phi^-\rangle_{C_1C_2},$$
$$|\Psi^+\rangle_{A_1A_2}|\Psi^+\rangle_{B_1B_2}|\Psi^-\rangle_{C_1C_2}, |\Psi^+\rangle_{A_1A_2}|\Psi^-\rangle_{B_1B_2}|\Psi^+\rangle_{C_1C_2}, |\Psi^-\rangle_{A_1A_2}|\Psi^+\rangle_{B_1B_2}|\Psi^+\rangle_{C_1C_2}, |\Psi^-\rangle_{A_1A_2}|\Psi^-\rangle_{B_1B_2}|\Psi^-\rangle_{C_1C_2}\} \to \mathbb{C}_1 \quad (4)$$

$$\{|\Psi^+\rangle_{A_1A_2}|\Phi^+\rangle_{B_1B_2}|\Phi^-\rangle_{C_1C_2}, |\Psi^+\rangle_{A_1A_2}|\Phi^-\rangle_{B_1B_2}|\Phi^+\rangle_{C_1C_2}, |\Psi^-\rangle_{A_1A_2}|\Phi^+\rangle_{B_1B_2}|\Phi^+\rangle_{C_1C_2}, |\Psi^-\rangle_{A_1A_2}|\Phi^-\rangle_{B_1B_2}|\Phi^-\rangle_{C_1C_2},$$
$$|\Phi^+\rangle_{A_1A_2}|\Psi^+\rangle_{B_1B_2}|\Psi^+\rangle_{C_1C_2}, |\Phi^+\rangle_{A_1A_2}|\Psi^-\rangle_{B_1B_2}|\Psi^-\rangle_{C_1C_2}, |\Phi^-\rangle_{A_1A_2}|\Psi^+\rangle_{B_1B_2}|\Psi^-\rangle_{C_1C_2}, |\Phi^-\rangle_{A_1A_2}|\Psi^-\rangle_{B_1B_2}|\Psi^+\rangle_{C_1C_2}\} \to \mathbb{C}_2 \quad (5)$$

$$\{|\Psi^+\rangle_{A_1A_2}|\Phi^+\rangle_{B_1B_2}|\Phi^-\rangle_{C_1C_2}, |\Psi^+\rangle_{A_1A_2}|\Phi^-\rangle_{B_1B_2}|\Phi^+\rangle_{C_1C_2}, |\Psi^-\rangle_{A_1A_2}|\Phi^+\rangle_{B_1B_2}|\Phi^+\rangle_{C_1C_2}, |\Psi^-\rangle_{A_1A_2}|\Phi^-\rangle_{B_1B_2}|\Phi^-\rangle_{C_1C_2},$$
$$|\Phi^+\rangle_{A_2A_2}|\Psi^+\rangle_{B_1B_2}|\Psi^-\rangle_{C_1C_2}, |\Phi^+\rangle_{A_1A_2}|\Psi^-\rangle_{B_1B_2}|\Psi^+\rangle_{C_1C_2}, |\Phi^-\rangle_{A_1A_2}|\Psi^+\rangle_{B_1B_2}|\Psi^+\rangle_{C_1C_2}, |\Phi^-\rangle_{A_1A_2}|\Psi^-\rangle_{B_1B_2}|\Psi^-\rangle_{C_1C_2}\} \to \mathbb{C}_3 \quad (6)$$

$$\{|\Phi^+\rangle_{A_1A_2}|\Psi^+\rangle_{B_1B_2}|\Phi^+\rangle_{C_1C_2}, |\Phi^+\rangle_{A_1A_2}|\Psi^-\rangle_{B_1B_2}|\Phi^-\rangle_{C_1C_2}, |\Phi^-\rangle_{A_1A_2}|\Psi^+\rangle_{B_1B_2}|\Phi^-\rangle_{C_1C_2}, |\Phi^-\rangle_{A_1A_2}|\Psi^-\rangle_{B_1B_2}|\Phi^+\rangle_{C_1C_2},$$
$$|\Psi^+\rangle_{A_1A_2}|\Phi^+\rangle_{B_1B_2}|\Psi^+\rangle_{C_1C_2}, |\Psi^+\rangle_{A_1A_2}|\Phi^-\rangle_{B_1B_2}|\Psi^-\rangle_{C_1C_2}, |\Psi^-\rangle_{A_1A_2}|\Phi^+\rangle_{B_1B_2}|\Psi^-\rangle_{C_1C_2}, |\Psi^-\rangle_{A_1A_2}|\Phi^-\rangle_{B_1B_2}|\Psi^+\rangle_{C_1C_2}\} \to \mathbb{C}_4 \quad (7)$$

$$\{|\Phi^+\rangle_{A_1A_2}|\Psi^+\rangle_{B_1B_2}|\Phi^-\rangle_{C_1C_2}, |\Phi^+\rangle_{A_1A_2}|\Psi^-\rangle_{B_1B_2}|\Phi^+\rangle_{C_1C_2}, |\Phi^-\rangle_{A_1A_2}|\Psi^+\rangle_{B_1B_2}|\Phi^+\rangle_{C_1C_2}, |\Phi^-\rangle_{A_1A_2}|\Psi^-\rangle_{B_1B_2}|\Phi^-\rangle_{C_1C_2},$$
$$|\Psi^+\rangle_{A_1A_2}|\Phi^+\rangle_{B_1B_2}|\Psi^-\rangle_{C_1C_2}, |\Psi^+\rangle_{A_1A_2}|\Phi^-\rangle_{B_1B_2}|\Psi^+\rangle_{C_1C_2}, |\Psi^-\rangle_{A_1A_2}|\Phi^+\rangle_{B_1B_2}|\Psi^+\rangle_{C_1C_2}, |\Psi^-\rangle_{A_1A_2}|\Phi^-\rangle_{B_1B_2}|\Psi^-\rangle_{C_1C_2}\} \to \mathbb{C}_5 \quad (8)$$

$$\{|\Psi^+\rangle_{A_1A_2}|\Psi^+\rangle_{B_1B_2}|\Phi^+\rangle_{C_1C_2}, |\Psi^+\rangle_{A_1A_2}|\Psi^-\rangle_{B_1B_2}|\Phi^-\rangle_{C_1C_2}, |\Psi^-\rangle_{A_1A_2}|\Psi^+\rangle_{B_1B_2}|\Phi^-\rangle_{C_1C_2}, |\Psi^-\rangle_{A_1A_2}|\Psi^-\rangle_{B_1B_2}|\Phi^+\rangle_{C_1C_2},$$
$$|\Phi^+\rangle_{A_1A_2}|\Phi^+\rangle_{B_1B_2}|\Psi^+\rangle_{C_1C_2}, |\Phi^+\rangle_{A_1A_2}|\Phi^-\rangle_{B_1B_2}|\Psi^-\rangle_{C_1C_2}, |\Phi^-\rangle_{A_1A_2}|\Phi^+\rangle_{B_1B_2}|\Psi^-\rangle_{C_1C_2}, |\Phi^-\rangle_{A_1A_2}|\Phi^-\rangle_{B_1B_2}|\Psi^+\rangle_{C_1C_2}\} \to \mathbb{C}_6 \quad (9)$$

$$\{|\Psi^+\rangle_{A_1A_2}|\Psi^+\rangle_{B_1B_2}|\Phi^-\rangle_{C_1C_2}, |\Psi^+\rangle_{A_1A_2}|\Psi^-\rangle_{B_1B_2}|\Phi^+\rangle_{C_1C_2}, |\Psi^-\rangle_{A_1A_2}|\Psi^+\rangle_{B_1B_2}|\Phi^+\rangle_{C_1C_2}, |\Psi^-\rangle_{A_1A_2}|\Psi^-\rangle_{B_1B_2}|\Phi^-\rangle_{C_1C_2},$$
$$|\Phi^+\rangle_{A_1A_2}|\Phi^+\rangle_{B_1B_2}|\Psi^-\rangle_{C_1C_2}, |\Phi^+\rangle_{A_1A_2}|\Phi^-\rangle_{B_1B_2}|\Psi^+\rangle_{C_1C_2}, |\Phi^-\rangle_{A_1A_2}|\Phi^+\rangle_{B_1B_2}|\Psi^+\rangle_{C_1C_2}, |\Phi^-\rangle_{A_1A_2}|\Phi^-\rangle_{B_1B_2}|\Psi^-\rangle_{C_1C_2}\} \to \mathbb{C}_7 \quad (10)$$

Moreover, after some similar deductions, we can obtain all the outcome collections composed of the different outcome combinations of $A_1A_2$, $B_1B_2$ and $C_1C_2$ after entanglement swapping between any two GHZ states, which is shown as Table 2[39].

Table 2. The outcome collections of entanglement swapping between any two GHZ states

| | $|\Psi_0\rangle_{A_2B_2C_2}$ | $|\Psi_1\rangle_{A_2B_2C_2}$ | $|\Psi_2\rangle_{A_2B_2C_2}$ | $|\Psi_3\rangle_{A_2B_2C_2}$ | $|\Psi_4\rangle_{A_2B_2C_2}$ | $|\Psi_5\rangle_{A_2B_2C_2}$ | $|\Psi_6\rangle_{A_2B_2C_2}$ | $|\Psi_7\rangle_{A_2B_2C_2}$ |
|---|---|---|---|---|---|---|---|---|
| $|\Psi_0\rangle_{A_1B_1C_1}$ | $\mathbb{C}_0$ | $\mathbb{C}_1$ | $\mathbb{C}_2$ | $\mathbb{C}_3$ | $\mathbb{C}_4$ | $\mathbb{C}_5$ | $\mathbb{C}_6$ | $\mathbb{C}_7$ |
| $|\Psi_1\rangle_{A_1B_1C_1}$ | $\mathbb{C}_1$ | $\mathbb{C}_0$ | $\mathbb{C}_3$ | $\mathbb{C}_2$ | $\mathbb{C}_5$ | $\mathbb{C}_4$ | $\mathbb{C}_7$ | $\mathbb{C}_6$ |
| $|\Psi_2\rangle_{A_1B_1C_1}$ | $\mathbb{C}_2$ | $\mathbb{C}_3$ | $\mathbb{C}_0$ | $\mathbb{C}_1$ | $\mathbb{C}_6$ | $\mathbb{C}_7$ | $\mathbb{C}_4$ | $\mathbb{C}_5$ |
| $|\Psi_3\rangle_{A_1B_1C_1}$ | $\mathbb{C}_3$ | $\mathbb{C}_2$ | $\mathbb{C}_1$ | $\mathbb{C}_0$ | $\mathbb{C}_7$ | $\mathbb{C}_6$ | $\mathbb{C}_5$ | $\mathbb{C}_4$ |
| $|\Psi_4\rangle_{A_1B_1C_1}$ | $\mathbb{C}_4$ | $\mathbb{C}_5$ | $\mathbb{C}_6$ | $\mathbb{C}_7$ | $\mathbb{C}_0$ | $\mathbb{C}_1$ | $\mathbb{C}_2$ | $\mathbb{C}_3$ |
| $|\Psi_5\rangle_{A_1B_1C_1}$ | $\mathbb{C}_5$ | $\mathbb{C}_4$ | $\mathbb{C}_7$ | $\mathbb{C}_6$ | $\mathbb{C}_1$ | $\mathbb{C}_0$ | $\mathbb{C}_3$ | $\mathbb{C}_2$ |
| $|\Psi_6\rangle_{A_1B_1C_1}$ | $\mathbb{C}_6$ | $\mathbb{C}_7$ | $\mathbb{C}_4$ | $\mathbb{C}_5$ | $\mathbb{C}_2$ | $\mathbb{C}_3$ | $\mathbb{C}_0$ | $\mathbb{C}_1$ |

| $\|\Psi_7\rangle_{A_1B_1C_1}$ | $\mathbb{C}_7$ | $\mathbb{C}_6$ | $\mathbb{C}_5$ | $\mathbb{C}_4$ | $\mathbb{C}_3$ | $\mathbb{C}_2$ | $\mathbb{C}_1$ | $\mathbb{C}_0$ |

## 3 The BQSDC protocol

We employ dense coding of GHZ states [38] and qubit transmission in blocks[4,7] to put forward a BQSDC protocol. Our protocol transmits three particle sequences composed by GHZ states from Alice to Bob one by one in three steps, which is identical to the transmission process of the multi-step QSDC proposed by Wang et.al[9]. Without loss of generality, suppose that Alice has $3N$ bits secret messages $\{(i_1,j_1,r_1)(i_2,j_2,r_2)\cdots(i_n,j_n,r_n)\cdots(i_N,j_N,r_N)\}$, and Bob has $3N$ bits secret messages $\{(k_1,l_1,t_1)(k_2,l_2,t_2)\cdots(k_n,l_n,t_n)\cdots(k_N,l_N,t_N)\}$, respectively, where $i_n,j_n,r_n,k_n,l_n,t_n \in \{0,1\}$, $n \in \{1,2,\cdots,N\}$. They agree on that the unitary operations are performed on the first and the second particles of a GHZ state for encoding. Our protocol is illustrated in detail now.

**Step 1: Preparation.** A sequence consisting of $N$ triples of binary numbers $(A_i, B_i, C_i)$ is randomly generated, where each triple corresponds to a GHZ state. Then the GHZ states are generated, where Alice uses one triple for generating two same GHZ states. The particles $A$, $B$ and $C$ from each GHZ state form three corresponding ordered particle sequences $S_A$, $S_B$ and $S_C$. That is, $S_A = \{A_1, A_2, \cdots, A_{2N}\}$, $S_B = \{B_1, B_2, \cdots, B_{2N}\}$ and $S_C = \{C_1, C_2, \cdots, C_{2N}\}$. Alice prepares another batch of sample GHZ states for the first eavesdropping check, and randomly inserts the sample particles $A$, $B$ and $C$ into the corresponding original $S_A$, $S_B$ and $S_C$. Consequently, $S_A$, $S_B$ and $S_C$ turn into three new sequences $S_A'$, $S_B'$ and $S_C'$. Alice sends $S_C'$ to Bob, and keeps $S_A'$ and $S_B'$ by herself.

**Step 2: The first eavesdropping check.** After Bob confirms Alice that he has received $S_C'$, Alice firstly tells Bob the positions of the sample particle $C$ in $S_C'$. Then, Bob randomly chooses $Z$-basis $(\{|0\rangle,|1\rangle\})$ or $X$-basis $(\{|+\rangle,|-\rangle\})$ to measure the sample particle $C$ in $S_C'$, and tells Alice his measurement basis and measurement results. Alice chooses the same measurement basis to measure the corresponding sample particle $A$ in $S_A'$ and the corresponding sample particle $B$ in $S_B'$, respectively. By comparing her measurement results with Bob's, Alice can know whether the quantum channel is safe or not. If the channel is safe, their measurement results should be highly correlated, according to the entanglement correlation from GHZ states. If Alice confirms that the channel is not safe, they abort the communication; otherwise, they enter Step 3.

**Step 3: Alice's encoding.** After getting rid of the sample particles, $S_A'$, $S_B'$ and $S_C'$ turn back into $S_A$, $S_B$ and $S_C$, respectively. Both Alice and Bob divide their own sequences into groups (a group contains two adjacent particles). That is, $(A_{2n-1}, A_{2n})$ forms a group in $S_A$, $(B_{2n-1}, B_{2n})$ forms a group in $S_B$, and $(C_{2n-1}, C_{2n})$ forms a group in $S_C$ $(n=1,2,\cdots,N)$. Alice and Bob agree on the following encoding rule: (i) they should perform their individual unitary operation on different particles of one group for encoding; (ii) as to each person, the position of the encoded particle $A$ should be identical to that of the encoded particle $B$. In order to encode her secret messages, Alice picks one of the eight composite unitary operations according to her secret messages $(i_n, j_n, r_n)$, where $U_{i_n j_n r_n} = U_{i_n j_n r_n}^A \otimes U_{i_n j_n r_n}^B$, and performs $U_{i_n j_n r_n}^A$ on particle $A_{2n-1}$ and $U_{i_n j_n r_n}^B$ on particle $B_{2n-1}$ $(n=1,2,\cdots,N)$. Then, Alice prepares a large number of sample single particles randomly in one of the four states $\{|0\rangle, |1\rangle, |+\rangle, |-\rangle\}$ for the second eavesdropping check, and randomly inserts these sample single particles into $S_B$. Accordingly, $S_B$ turns into a new sequence $S_B''$. Afterward, Alice sends $S_B''$ to Bob.

**Step 4: The second eavesdropping check.** After Bob confirms Alice that he has received $S_B''$, Alice publishes the positions and the preparation basis of sample single particles to Bob. Then, Bob measures sample single particles in the same basis as the preparation basis of Alice, and announces his measurement results to Alice. By comparing the initial states of sample single particles with Bob's measurement results, Alice can know whether the quantum channel is safe or not. If Alice confirms that the channel is not safe, they abort the communication; otherwise, they enter Step 5.

**Step 5: The third eavesdropping check.** Alice prepares a large number of sample single particles randomly in one of the four states $\{|0\rangle, |1\rangle, |+\rangle, |-\rangle\}$ for the third eavesdropping check, and randomly inserts these sample single particles into $S_A$. Accordingly, $S_A$ turns into a new sequence $S_A''$. Then, Alice sends $S_A''$ to Bob. After Bob confirms Alice that he has received $S_A''$,

Alice publishes the positions and the preparation basis of sample single particles to Bob. Then, Bob measures sample single particles in the same basis as the preparation basis of Alice, and announces his measurement results to Alice. By comparing the initial states of sample single particles with Bob's measurement results, Alice can know whether the quantum channel is safe or not. If Alice confirms that the channel is not safe, they abort the communication; otherwise, they enter Step 6.

**Step 6: Bob's encoding.** After getting rid of the sample single particles, $S_B^{"}$ and $S_A^{"}$ turn back into $S_B$ and $S_A$ again, respectively. Having $S_A$, $S_B$ and $S_C$ in hand, Bob picks up one particle from each sequence in order, and stores each two adjacent GHZ states as a group. In other words, group $n$ contains two GHZ states $\{(U_{i_n j_n r_n}^A A_{2n-1}, U_{i_n j_n r_n}^B B_{2n-1}, C_{2n-1}), (A_{2n}, B_{2n}, C_{2n})\}$ $(n=1,2,\cdots,N)$. Then, Bob performs GHZ-basis measurement on $(A_{2n}, B_{2n}, C_{2n})$ at first, so he will know the initial state of group $n$ prepared by Alice in Step 1. According to his GHZ-basis measurement outcome, Bob reproduces a new $(A_{2n}, B_{2n}, C_{2n})$ in which state measurement is not performed. In order to encode his secret messages, Bob picks one of the eight composite unitary operations according to his secret messages $(k_n, l_n, t_n)$, where $U_{k_n l_n t_n} = U_{k_n l_n t_n}^A \otimes U_{k_n l_n t_n}^B$, and performs $U_{k_n l_n t_n}^A$ on the new particle $A_{2n}$ and $U_{k_n l_n t_n}^B$ on the new particle $B_{2n}$. Consequently, group $n$ turns into $\{(U_{i_n j_n r_n}^A A_{2n-1}, U_{i_n j_n r_n}^B B_{2n-1}, C_{2n-1}), (U_{k_n l_n t_n}^A A_{2n}, U_{k_n l_n t_n}^B B_{2n}, C_{2n})\}$.

**Step 7: Bidirectional communication.** Bob performs Bell-basis measurements on $(U_{i_n j_n r_n}^A A_{2n-1}, U_{k_n l_n t_n}^A A_{2n})$, $(U_{i_n j_n r_n}^B B_{2n-1}, U_{k_n l_n t_n}^B B_{2n})$ and $(C_{2n-1}, C_{2n})$. According to formulas (3)-(10), Bob can know which outcome collection $\{(U_{i_n j_n r_n}^A A_{2n-1}, U_{k_n l_n t_n}^A A_{2n}), (U_{i_n j_n r_n}^B B_{2n-1}, U_{k_n l_n t_n}^B B_{2n}), (C_{2n-1}, C_{2n})\}$ belongs to. Then, Bob announces Alice the outcome collection through the classical channel. According to the outcome collection from Bob's announcement, Alice can infer the eight possible kinds of states about $\{(U_{i_n j_n r_n}^A A_{2n-1}, U_{i_n j_n r_n}^B B_{2n-1}, C_{2n-1}), (U_{k_n l_n t_n}^A A_{2n}, U_{k_n l_n t_n}^B B_{2n}, C_{2n})\}$ from Table 2. Moreover, since Alice knows the initial state of group $n$ from her preparation in Step 1, according to her composite unitary operation $U_{i_n j_n r_n}$, Alice is able to know the state of $(U_{k_n l_n t_n}^A A_{2n}, U_{k_n l_n t_n}^B B_{2n}, C_{2n})$ from Table 2. Consequently, Alice can deduce Bob's composite unitary operation $U_{k_n l_n t_n}$ from Table 1. Accordingly, Alice can infer Bob's secret messages $(k_n, l_n, t_n)$. Similarly, according to the outcome collection $\{(U_{i_n j_n r_n}^A A_{2n-1}, U_{k_n l_n t_n}^A A_{2n}), (U_{i_n j_n r_n}^B B_{2n-1}, U_{k_n l_n t_n}^B B_{2n}), (C_{2n-1}, C_{2n})\}$ belongs to, Bob can also know the eight possible kinds of states about $\{(U_{i_n j_n r_n}^A A_{2n-1}, U_{i_n j_n r_n}^B B_{2n-1}, C_{2n-1}), (U_{k_n l_n t_n}^A A_{2n}, U_{k_n l_n t_n}^B B_{2n}, C_{2n})\}$ from Table 2. Since Bob knows the initial state of group $n$ from his GHZ-basis measurement on $(A_{2n}, B_{2n}, C_{2n})$, according to his composite unitary operation $U_{k_n l_n t_n}$, Bob is able to know the state of $(U_{i_n j_n r_n}^A A_{2n-1}, U_{i_n j_n r_n}^B B_{2n-1}, C_{2n-1})$ from Table 2. Consequently, Bob can infer Alice's secret messages $(i_n, j_n, r_n)$ from Table 1. Apparently, $(A_{2n}, B_{2n}, C_{2n})$ acts as a shared secret GHZ state between Alice and Bob. It plays two roles: on one hand, it lets Bob know the prepared initial state; on the other hand, it is used for encoding Bob's secret messages.

Without loss of generality, we take the first two adjacent GHZ states $(A_1, B_1, C_1)$ and $(A_2, B_2, C_2)$ for an example to further explain the bidirectional communication process here. Suppose that $(i_1, j_1, r_1)$ and $(k_1, l_1, t_1)$ are **010** and **101**, respectively. Moreover, suppose that the initial state of both $(A_1, B_1, C_1)$ and $(A_2, B_2, C_2)$ is $|\Psi_0\rangle$. Therefore, $(A_1, B_1, C_1)$ and $(A_2, B_2, C_2)$ evolve as follows:

$$\left. \begin{array}{l} |\Psi_0\rangle_{A_1 B_1 C_1} \Rightarrow i\sigma_y^{A_1} \otimes \sigma_z^{B_1} |\Psi_0\rangle_{A_1 B_1 C_1} = |\Psi_2\rangle_{A_1 B_1 C_1} \\ |\Psi_0\rangle_{A_2 B_2 C_2} \Rightarrow \sigma_z^{A_2} \otimes \sigma_x^{B_2} |\Psi_0\rangle_{A_2 B_2 C_2} = |\Psi_5\rangle_{A_2 B_2 C_2} \end{array} \right\} \Rightarrow |\Psi_2\rangle_{A_1 B_1 C_1} \otimes |\Psi_5\rangle_{A_2 B_2 C_2} \to \mathbb{C}_7 \qquad (11)$$

The $\mathbb{C}_7$ is the outcome collection which the entanglement swapping outcome of $|\Psi_2\rangle_{A_1 B_1 C_1} \otimes |\Psi_5\rangle_{A_2 B_2 C_2}$ belongs to, and is published by Bob. According to three known information: $\mathbb{C}_7$, the prepared initial state $|\Psi_0\rangle$ and $U_2(i\sigma_y^{A_1} \otimes \sigma_z^{B_1})$, Alice can infer from Tables 1 and 2 that Bob's secret messages $(k_1, l_1, t_1)$ are **101**. Similarly, Bob can infer from Tables 1 and 2 that Alice's secret messages $(i_1, j_1, r_1)$ are **010**, according to $\mathbb{C}_7$, the initial state $|\Psi_0\rangle$ known from his GHZ-basis measurement

on $(A_2, B_2, C_2)$ and $U_5 \left(\sigma_z^{A_2} \otimes \sigma_x^{B_2}\right)$. Therefore, they successfully exchange their own secret messages.

## 4 Analysis and Discussions
### 4.1 Security analysis on eavesdropping checks

Although the transmission process of our protocol is identical to that of the multi-step QSDC[9], the eavesdropping checking methods of our protocol are not completely identical to those of the multi-step QSDC[9]. Our protocol implements eavesdropping check three times in total. The first eavesdropping check uses the entanglement correlation from GHZ states to check whether the sequence $S_C^{'}$ is transmitted securely, which has been also used to check the security of a transmitted sequence in Refs.[9,20,21,23,39]. (I) The intercept-resend attack. Eve intercepts the sequence $S_C^{'}$, then sends her fake sequence prepared in advance instead of it to Bob. Since the original entanglement correlations between particle $A$, particle $B$ and particle $C$ have been destroyed, Eve can be detected when the first eavesdropping check is implemented. Without loss of generality, suppose that the initial GHZ state is $|\Psi_0\rangle_{ABC}$. If there is no Eve in the line, the measurement result of Alice and Bob is $|000\rangle$ or $|111\rangle$ ($|+\rangle|+\rangle|+\rangle, |-\rangle|-\rangle|+\rangle, |+\rangle|-\rangle|-\rangle$ or $|-\rangle|+\rangle|-\rangle$). If Eve is in the line, she intercepts the particle $C$ and sends the particle $c$ to Bob. If the particle $c$ is prepared by Eve in state $|0\rangle_c$, given that Alice and Bob choose $Z$-basis ($X$-basis), their measurement result is $|000\rangle_{ABc}$ or $|110\rangle_{ABc}$ ($|opq\rangle_{ABc}, o, p, q = +, -$). Thus, Eve's eavesdropping will be detected with probability $50\%$ ($50\%$). If the particle $c$ is prepared in state $|1\rangle_c$, Eve's eavesdropping will also be detected with probability $50\%$ ($50\%$). If the particle $c$ is prepared in state $|+\rangle_c$, Eve's eavesdropping will be detected with probability $75\%$ ($50\%$). If the particle $c$ is prepared in state $|-\rangle_c$, Eve's eavesdropping will also be detected with probability $75\%$ ($50\%$). (II) The measurement-resend attack. After intercepting the sequence $S_C^{'}$, Eve measures it at first. Then, she resends it to Bob. Since the measuring basis that Alice and Bob choose are not always consistent with that of Eve, this eavesdropping attack can be detected. Without loss of generality, suppose that the initial GHZ state is $|\Psi_0\rangle_{ABC}$. Eve intercepts the particle $C$, measures it in $Z$-basis or $X$-basis, and resends its measurement result to Bob. In the first case, Eve performs $Z$-basis measurement. The state of the whole system will collapse to $|000\rangle$ or $|111\rangle$ each with probability $1/2$. Take the state to be $|000\rangle_{ABC}$ for example. Accordingly, Eve resends $|0\rangle_C$ to Bob. Then, if Bob performs $Z$-basis measurement to check security, no error will be introduced by Eve. If Bob performs $X$-basis measurement, the state collapses to $|opq\rangle_{ABC} (o, p, q = +, -)$ each with probability $1/8$. Thus, the error rate introduced by Eve will be $50\%$. Therefore, the total error rate in this case is $25\%$. In the second case, Eve performs $X$-basis measurement. The state of the whole system will collapse to $|+\rangle_A|+\rangle_B|+\rangle_C, |+\rangle_A|-\rangle_B|-\rangle_C, |-\rangle_A|+\rangle_B|-\rangle_C$ or $|-\rangle_A|-\rangle_B|+\rangle_C$ each with probability $1/4$. Take the state to be $|+\rangle_A|+\rangle_B|+\rangle_C$ for example. Accordingly, Eve resends $|+\rangle_C$ to Bob. Then, if Bob performs $Z$-basis measurement to check security, the state collapses to $|opq\rangle_{ABC} (o, p, q = 0, 1)$ each with probability $1/8$. Thus, the error rate introduced by Eve will be $75\%$. If Bob performs $X$-basis measurement, no error will be introduced by Eve. Therefore, the total error rate in this case is $37.5\%$.[39] (III) The entanglement-and-measurement attack. Eve may steal partial information by entangling her auxiliary particle (prepared in the state $|\varepsilon_i\rangle_e$) with the particle (assumed to be in the state of $|i\rangle, i \in \{0,1\}$) in sequence $S_C^{'}$. Then it follows,

$$\hat{E}|i\rangle \otimes |\varepsilon_i\rangle_e = \alpha|i\rangle|\varepsilon_i\rangle_e + \beta|i\oplus 1\rangle|\varepsilon_{i\oplus 1}\rangle_e \tag{12}$$

where $\hat{E}$ is Eve's unitary operation, $|\alpha|^2 + |\beta|^2 = 1$ and $_e\langle\varepsilon_i|\varepsilon_{i\oplus 1}\rangle_e = 0$. Apparently, with a probability of $|\beta|^2$, Eve will be detected when Alice and Bob implement the first eavesdropping check under $Z$-basis. [6-7]

The second eavesdropping check is to check the transmission of sequence $S_B^{''}$, while the third eavesdropping check is to check the transmission of sequence $S_A^{''}$. In our protocol, the checking method used in the second eavesdropping check is the same as that of the third eavesdropping check, which is derived from the idea of the BB84 QKD protocol[1]. This checking method has also been used in Refs.[21,22,25,26,31,35]. (I) The intercept-resend attack. Eve intercepts the sequence $S_B^{''}$ ($S_A^{''}$), then sends her fake sequence prepared in advance instead of it to Bob. Eve can be detected with probability of $50\%$, since Bob's measurement results on the fake sequence are not always identical with the genuine ones. [25,35] (II) The measurement-resend attack. After intercepting the sequence $S_B^{''}$ ($S_A^{''}$), Eve measures it at first. Then, she resends it to Bob. Since Eve's measurement basis is not always consistent with Alice's preparation basis for sample single particle, this eavesdropping attack can be detected with probability $25\%$.[25,35] (III)The entanglement-and-measurement attack. Eve may steal partial information by entangling her auxiliary particle (prepared in the state $|\varepsilon_i\rangle_e$) with the particle (assumed to be in the state of $|i\rangle$, $i \in \{0,1\}$) in sequence $S_B''$ ($S_A''$). Apparently, according to formula (12), Eve will be detected with probability $|\beta|^2$ when Alice's preparation basis for the sample single particle is $Z$-basis. [25,35]

### 4.2 The information leakage problem

Without loss of generality, we also take the first two adjacent GHZ states $(A_1, B_1, C_1)$ and $(A_2, B_2, C_2)$ for an example to analyze the information leakage problem. With the help of a shared secret quantum state $(A_2, B_2, C_2)$, Alice and Bob can exchange their secret messages simultaneously. However, Eve knows nothing about the initial state of $(A_2, B_2, C_2)$. The only thing she can do is a pure guess. Therefore, Bob' announcement on the outcome collection of $\{(U^A_{i_1 j_1 r_1} A_1, U^A_{k_1 l_1 t_1} A_2), (U^B_{i_1 j_1 r_1} B_1, U^B_{k_1 l_1 t_1} B_2), (C_1, C_2)\}$ belongs to totally 64 kinds of composite unitary operation combinations from Alice and Bob to Eve. It is equivalent to say that the quantum channel contains $-\sum_{i=1}^{64} p_i \log_2 p_i = -64 \times \frac{1}{64} \log_2 \frac{1}{64} = 6$ bits secret messages for Eve from the viewpoint of information theory, which are equal to the total amount of secret messages from both Alice and Bob. So, no information leakage has happened. It is the shared secret quantum state $(A_2, B_2, C_2)$ that helps to avoid the information leakage problem. In addition, in our protocol, six bits secret messages are transmitted by six qubits, which means that our protocol has the highest capacity in Holevo bound[29,40]. This also indicates that no information leakage has happened in our protocol.

### 4.3 Capacity and efficiency

In our protocol, both Alice and Bob exchange their individual three bits secret messages per round communication, therefore the capacity of our protocol is six bits per round communication. In addition, we consider the information-theoretical efficiency defined by Cabello[41], i.e., $\eta = b_s / (q_t + b_t)$, where $b_s$, $q_t$ and $b_t$ are the expected secret bits received, the qubit used and the classical bits exchanged between Alice and Bob. In our protocol, each two adjacent GHZ states $(A_{2n-1}, B_{2n-1}, C_{2n-1})$ and $(A_{2n}, B_{2n}, C_{2n})$ is used for exchanging three bits from Alice and three bits from Bob, where three classical bits are needed for Bob to announce Alice the outcome collection $\{(U^A_{i_n j_n r_n} A_{2n-1}, U^A_{k_n l_n t_n} A_{2n}), (U^B_{i_n j_n r_n} B_{2n-1}, U^B_{k_n l_n t_n} B_{2n}), (C_{2n-1}, C_{2n})\}$ belongs to through the classical channel. Therefore, the information-theoretical efficiency of our protocol is $\eta = (3+3)/(6+3) = 66.7\%$.

### 4.4 Comparison of previous BQSDC protocols

Our protocol can totally transmit six bits per round communication without information leakage. However, in Zhang' protocol [15-16], Nguyen' protocol[18], Man's protocol[19], Chen's protocol[25], Shan's protocol[30] and Ye's second protocol in Ref.[35], four bits secret messages are transmitted per round communication with two bits leaked out; in Jin's protocol [20], Man's protocol[21], Man's protocol[23] and Ye's first protocol in Ref.[35], four bits are transmitted per round communication with three bits leaked out; in Man's improved version[24], three bits are transmitted per round communication with two bits leaked out; in Ji's protocol[22] and Yang's protocol[26], two bits are transmitted per round communication with one bit leaked out. In addition, in Shi's protocol[31] and Gao's protocols[34], four bits are transmitted per round communication without information leakage; in Shi's protocol[32], two bits are transmitted per round communication without information leakage; in Shi's protocol[33], three bits are transmitted per round communication without information leakage.

On the other hand, we compare the information-theoretical efficiency of our protocol with those of protocols in Refs.[31-34], since all of these protocols avoid the information leakage problem. As we know, in Shi's protocol[31], Alice and Bob exchange their individual two bit secret messages per round communication by using four qubits and two bit classical information, thus its efficiency is $\eta = (2+2)/(4+2) = 66.7\%$; in Shi's protocol[32], Alice and Bob exchange their individual one bit per round communication by using two qubits and one bit classical information, thus its efficiency is $\eta = (1+1)/(2+1) = 66.7\%$; in Shi's protocol[33], Alice and Bob exchange two bits and one bit per round communication respectively by using three qubits and one bit classical information, thus its efficiency is $\eta = (1+2)/(3+1) = 75\%$; in Gao's protocols[34], Alice and Bob also exchange their individual two bits per round communication by using four qubits and two bit classical information, thus its efficiency is also $\eta = (2+2)/(4+2) = 66.7\%$. Therefore, the efficiency of our protocol is equal to those of protocols in Refs.[31,32,34], and is smaller than that of protocol in Ref.[33].

To sum up, compared with those previous BQSDC protocols[15,16,18-26,30-35], the advantage of our protocol lies in having the following two characters simultaneously: on one hand, it can overcome the information leakage problem; one the other hand, its capacity is as high as six bits per round communication.

## 5 Conclusions

It is easy to know that our protocol is based on entanglement swapping between any two GHZ states. The key to design our protocol lies in the consistency among two initial GHZ states and the outcome collection composed of the different outcome combinations of $A_1A_2$, $B_1B_2$ and $C_1C_2$ after entanglement swapping, which is shown in Table 2. It can be verified that the consistency among two initial Bell states and the outcome collection composed of the different outcome combinations of $A_1A_2$ and $B_1B_2$ after entanglement swapping does also exist. Therefore, it is feasible to design a similar BQSDC protocol based on entanglement swapping between any two Bell states. Apparently, the information-theoretical efficiency in this case is $\eta = (2+2)/(4+2) = 66.7\%$, equal to that of our GHZ state version described as above. However, its capacity is only four bits per round communication, smaller than that of our GHZ state version. Therefore, it can be concluded that compared with the similar protocol using Bell states as the quantum resource, our GHZ state version takes advantage in the communication capacity per round.

To sum up, we have put forward a large payload BQSDC protocol without information leakage using entanglement swapping between any two GHZ states. Our protocol avoids the information leakage problem using a shared secret GHZ state, which plays two roles in the bidirectional communication process: on one hand, it lets Bob know the prepared initial state; on the other hand, it is used for encoding Bob's secret messages. Compared with those previous BQSDC protocols, the advantage of our protocol lies in having the following two characters simultaneously: on one hand, it overcomes the information leakage problem; one the other hand, its capacity is as high as six bits per round communication.

**Acknowledgements**

Funding by the National Natural Science Foundation of China (Grant No. 11375152), and the Natural Science Foundation of Zhejiang Province (Grant No. LQ12F02012) is gratefully acknowledged.